\documentclass[11pt]{article}
\usepackage{newpasp,epsf}
\def\edcomment#1{\iffalse\marginpar{\raggedright\sl#1\/}\else\relax\fi}
\reversemarginpar
\newcommand{\gtsim}{\mbox
{{\raisebox{-0.4ex}{$\stackrel{>}{{\scriptstyle\sim}}$}}}}

\begin{document}
\title{3C radio sources as they've never been seen before} 
 \author{Katherine M.\ Blundell}
\affil{University of Oxford, Astrophysics, Keble Road, Oxford, OX1 3RH, UK}
\author{Namir E.\ Kassim}
\affil{NRL, Code 7213, Washington DC, 20375-5351, USA}
\author{Rick A.\ Perley}
\affil{NRAO, P.O.\ Box 0, Socorro NM, 87801-0387, USA }

\begin{abstract}
Low-radio-frequency observations played a remarkable role in the early
days of radio astronomy; however, in the subsequent three or four decades
their usefulness has largely been in terms of the {\sl finding-frequency}
of surveys.  Recent technical innovation at the VLA has meant that {\sl
spatially well-resolved} imaging at low frequencies is now possible.  Such
imaging is essential to understanding the relationship between the hotspot
and lobe emission in classical double radio sources, for example.  We here
present new images of 3C radio sources at 74\,MHz and 330\,MHz and discuss
their implications.
\end{abstract}

\section{Introduction}

Low-frequency radio surveys play a key role in selecting samples of radio
sources, dominated by optically thin synchrotron emission, which are free
of orientation biases.  Examples of such samples, which have been pivotal
in advancing our understanding of the nature of radio sources, are the
celebrated 3C sample (revised by Laing, Riley and Longair 1983) and very
recently the much fainter 7C sample (Rawlings et al.\ 1998).  Following
recent technical innovation at the VLA, it is now possible to make
spatially well-resolved images of radio sources at low frequencies: at
74\,MHz, images can now routinely be made with an angular resolution of
25$^{\prime\prime}$.  It is in this low-frequency regime that models of
the energy supply to the lobes from the hotspots in classical doubles may
be tested, and where their energy budgets may be investigated, for example
by the detection or otherwise of steep-spectrum halos surrounding these
objects.

\section{3C\,84 and its halo}
\begin{figure}[!h]
\plottwo{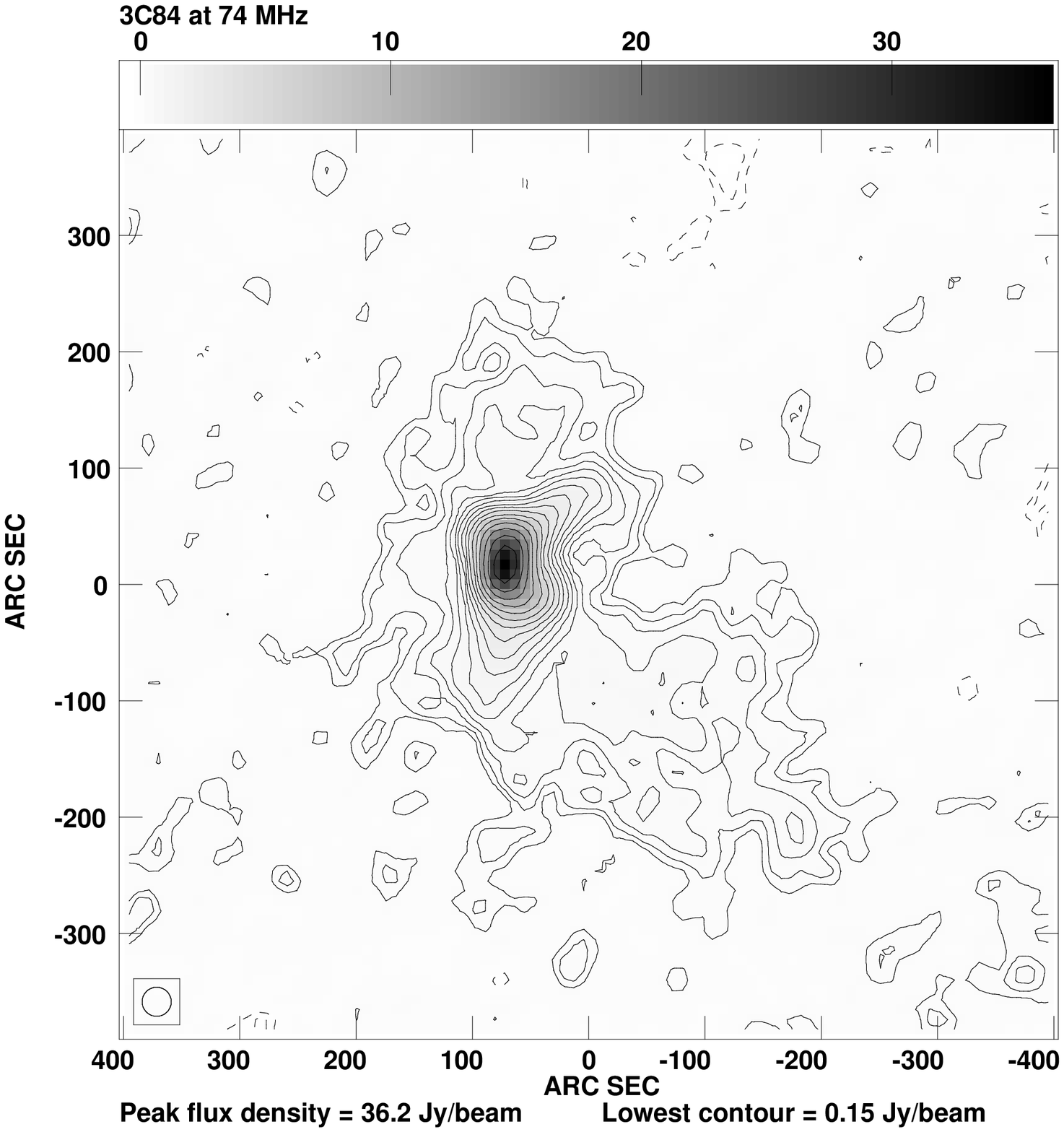}{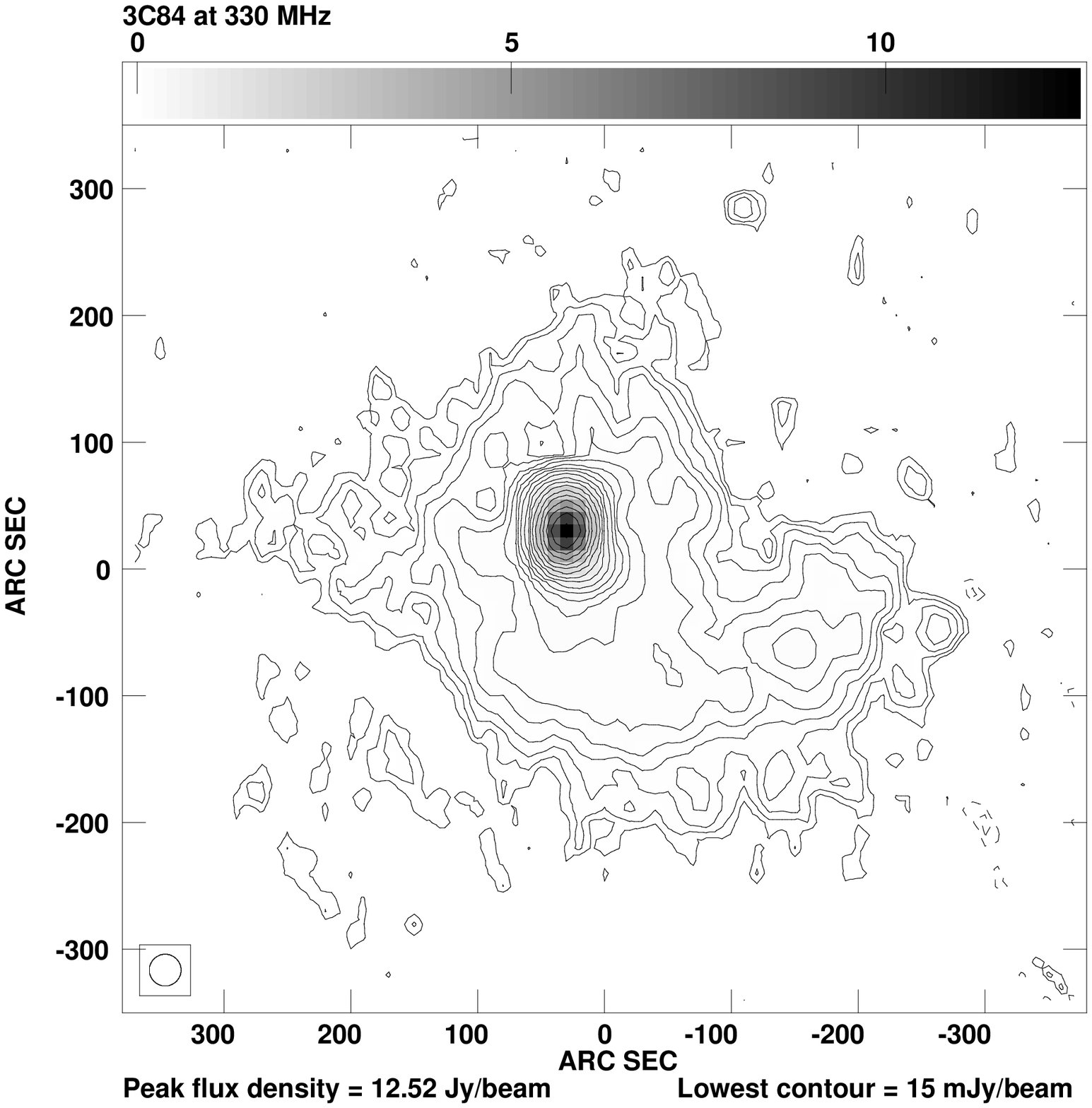}
\caption{Images of 3C84 at 74\,MHz and 330\,MHz with 25$^{\prime\prime}$
resolution. }
\end{figure}

The 74\,MHz image (Fig.\ 1, left) shows a low surface brightness halo
surrounding 3C\,84.  The signal to noise of the image is insufficient to
reveal whether the halo has a definite boundary, though this does appear
to be hinted at in the 330\,MHz image (see Fig.\ 1 right, and Burns et al
1992) and at 1.4\,GHz [Ger de Bruyn, priv.\ comm.].  The spectrum of the
halo is not particularly steep, averaging $\alpha^{74}_{330} \sim 1.1$
(defined such that the flux density $S_{\nu}$ at frequency $\nu$ is given
by $S_\nu \propto \nu^{-\alpha}$), with (e.g.\ the region
200$^{\prime\prime}$ west and 100$^{\prime\prime}$ south of the core) in
places a spectral index as flat as 0.7.  However, curious new features are
seen (only) in the 74\,MHz image: protrusions apparently emanating from
the core region (at 2 o'clock and 6 o'clock) have very steep spectra
($\alpha\ \gtsim\ 2)$.  It is possible that these could be outflows from
the core, or merely static structures with very strange spectra.  A full
analysis of these images and those of the other objects observed will be
presented in Kassim, Perley \& Blundell (in prep).

\section{3C\,129 and its twin tails}
While in the inner few arcmin this radio source is a wide-angle tailed
source with oppositely directed jets close to the core (see Rudnick \&
Burns 1981) on larger scales the jets follow each other quite closely and
have the appearance of a narrow-angle tailed source (see Fig.\ 2 and
Kassim et al.\ 1993).  The integrated spectral index of both the entire
source, and of just the extended tails, between 74\,MHz and 327\,MHz is
$\alpha \sim 1.1$.
\begin{figure}[!h]
\plotone{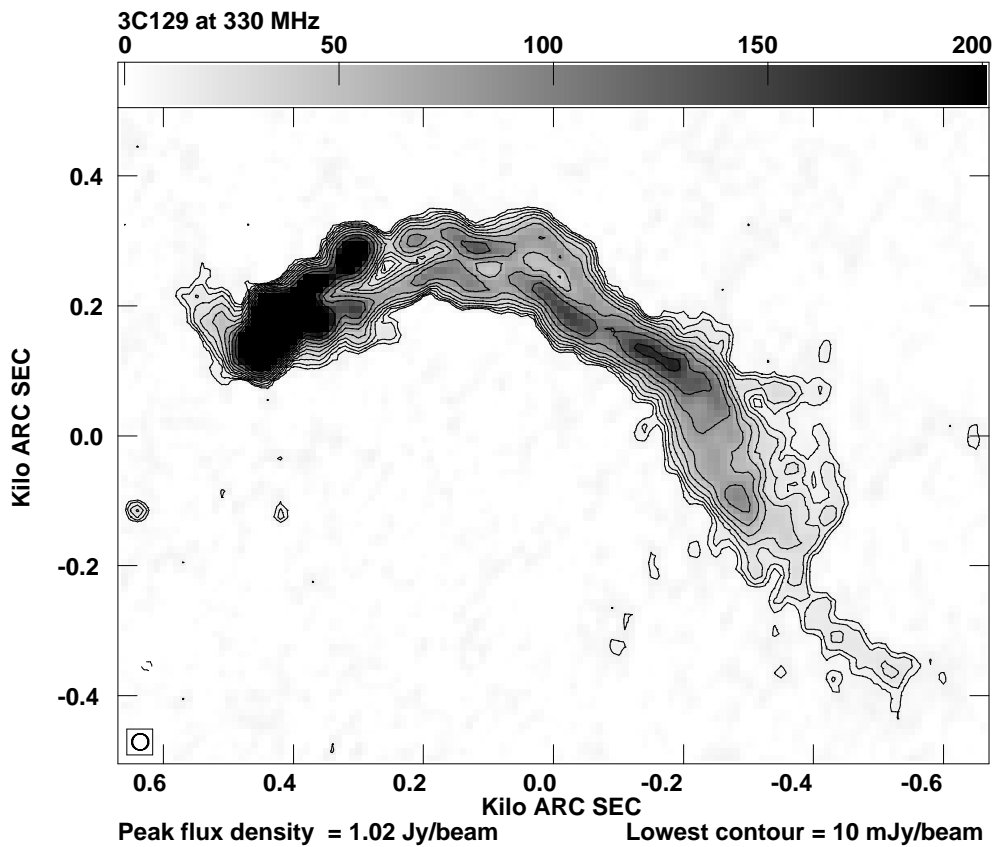}
\caption{ Image of 3C\,129 at 330\,MHz, showing the \gtsim\ 500 kpc extent
of its tails.  The greyscale is saturated above 0.2\,Jy/beam in order to
emphasise the twin tails which appear to persist as two distinct entities
over much of this distance.  }
\end{figure}

\section{3C\,219 and other classical doubles at low-frequency}
\begin{figure}[!h]
\plottwo{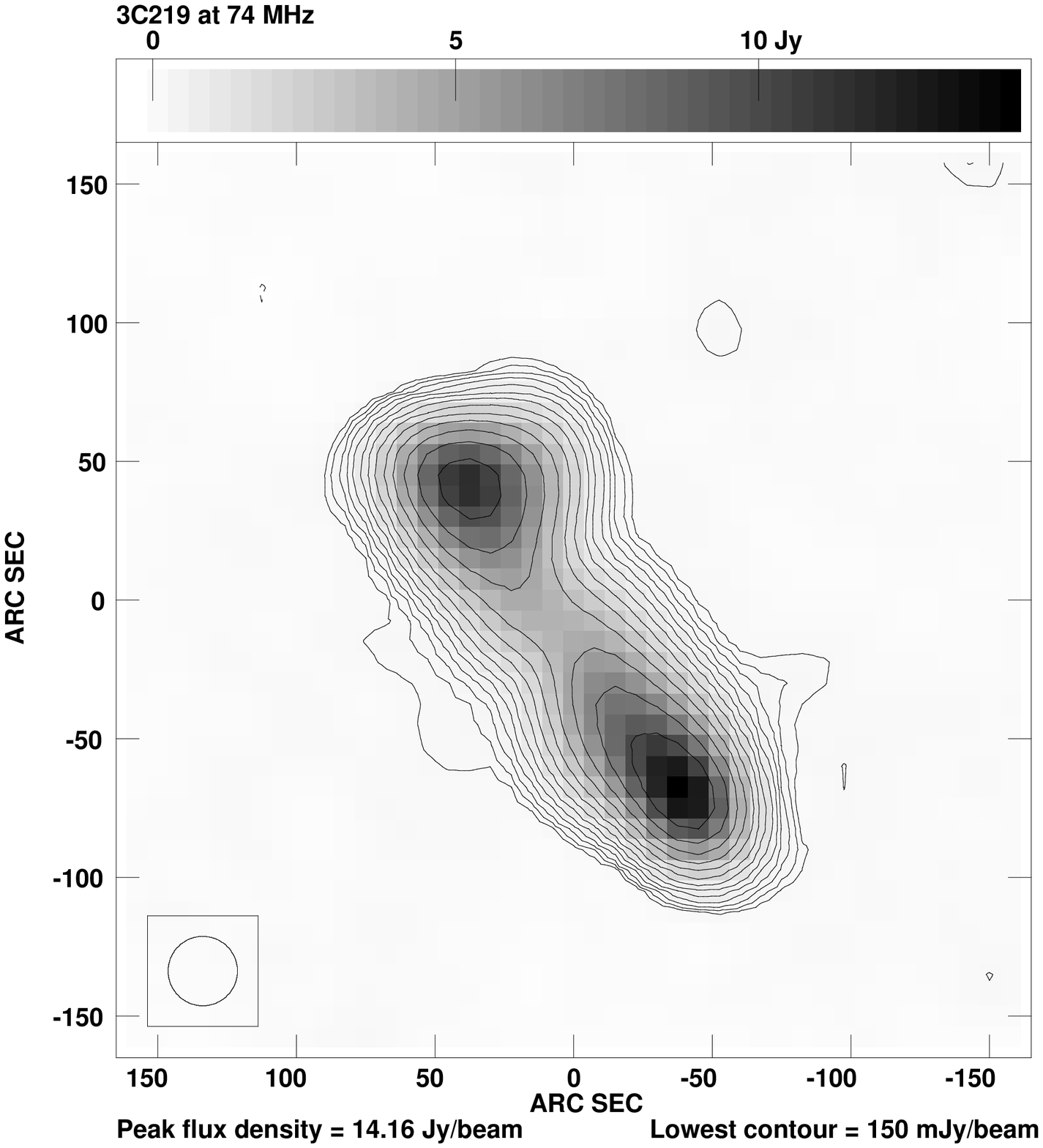}{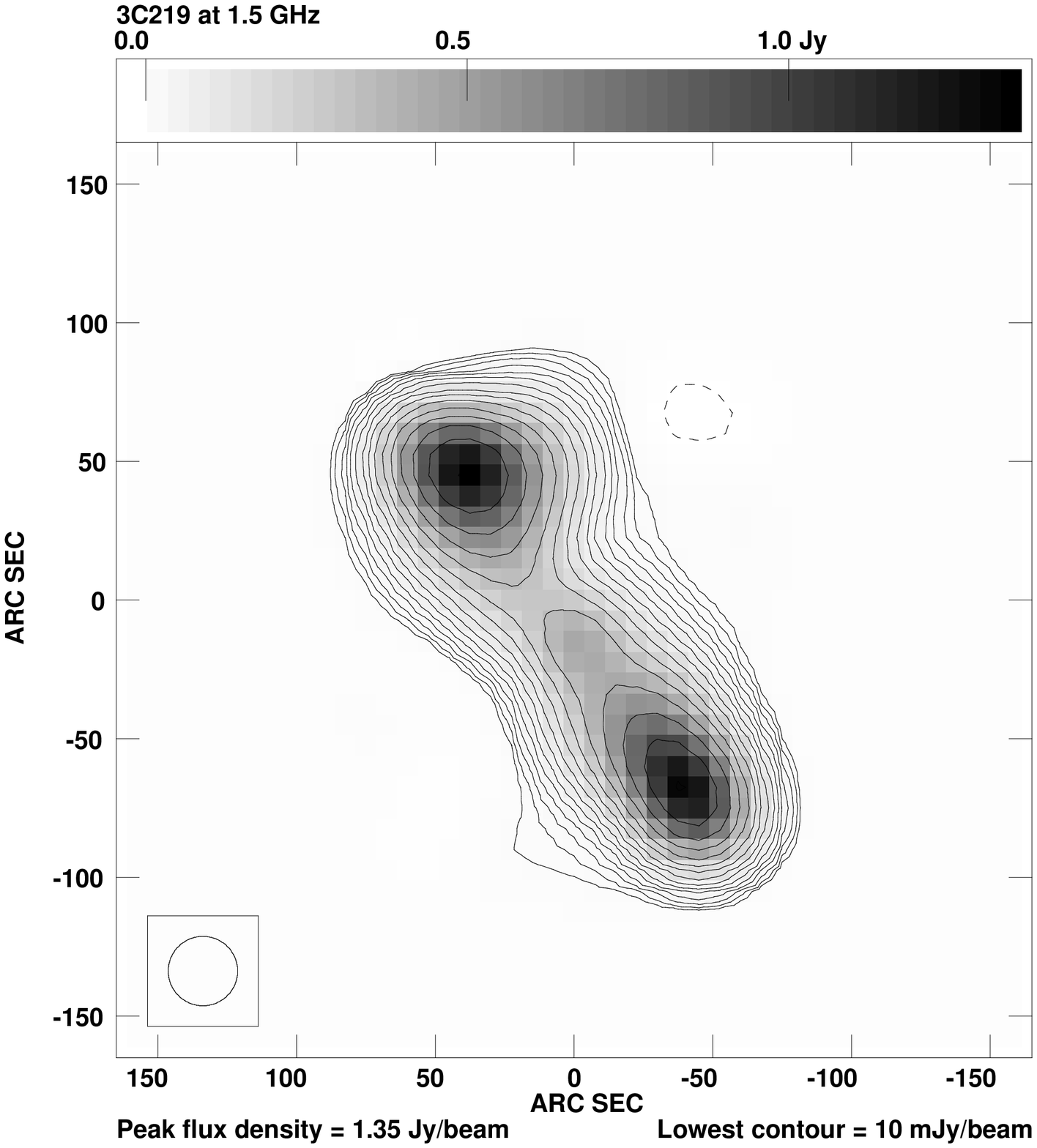}
\caption{Images of 3C\,219 at 74\,MHz and 1.5\,GHz with
25$^{\prime\prime}$ resolution.  These images, especially their
silhouettes, are more remarkable for their similarities than for their
differences.  } 
\end{figure}
Fig.\ 3 shows that the images of the classical double 3C\,219 at
74\,MHz and at 1.5\,GHz are more remarkable for their similarities
than for their differences.  Just as in the cases of 3C\,98 and
3C\,390.3 we presented in Blundell et al.\ (1999b) there is no
evidence of any extended emission at low-frequency which is not
already seen at GHz frequencies; this appears to be the case for all
the classical doubles imaged to date.  Jenkins \& Scheuer (1976)
pointed out that if synchrotron cooling played a part in determining
the spectral shapes of lobes, then {\sl lobes should be observed to
extend further at low frequency than at high frequency}.  Our images
thus suggest that the Lorentz factor particles responsible for the
74\,MHz emission are entirely co-spatial with those responsible for
the 1.4\,GHz emission.  

In a recent paper, Blundell \& Rawlings (2000) discussed a
contribution to spectral steepening along the lobe from the hotspot to
the core which is separate from the traditionally assumed simple
synchrotron cooling picture.  This contribution is particularly
important in explaining spectral gradients measured a decade below
fitted break frequencies, for example that in Cygnus A by Kassim et
al.\ (1996).  This comes from a gradient in magnetic field causing
different parts of the underlying curved energy distribution to be
`illuminated' (see figure 4 in Blundell \& Rawlings 2000) in the
different regions along the lobe.

We conclude by noting that low-frequency observations of 3C radio
sources have yet to reveal extended emission associated with a
classical double source beyond that seen at GHz frequencies.  Halos
around the rather more amorphous types of object are not uncommon and
in the case of 3C\,84 close to the core discrete regions with very
steep spectra ($\alpha\ \gtsim\ 2$) have been discovered at 74\,MHz.
A full analysis of these data will appear in a forthcoming paper.

It is a pleasure to thank the conference organisers for the splendid way
in which they organised the Pune meeting.  K.M.B.\ thanks the Royal
Commission for the Exhibition of 1851 for a Research Fellowship.  Basic
research in radio astronomy at the US Naval Research Laboratory is
supported by the US Office of Naval Research.  The VLA is a facility of
the NRAO operated by Associated Universities, Inc., under co-operative
agreement with the NSF.

\end{document}